\documentclass[aps, amsmath, amssymb, aps, pre, twocolumn,groupedaddress,superscriptaddress,nofootinbib,longbibliography]{revtex4-1}

\usepackage{balance}
\usepackage{amsfonts, amsmath, amssymb, xfrac}
\usepackage{amsthm}
\usepackage{enumitem}
\setlist{noitemsep,topsep=0pt,parsep=0pt,partopsep=0pt}
\usepackage{array}
\usepackage[english]{babel}

\usepackage{graphics}
\usepackage{graphicx}
\usepackage[config, labelfont={sf,bf}, textfont=sf]{caption,subfig}
\usepackage{bm}
\usepackage{lineno,hyperref}
\modulolinenumbers[5]

\theoremstyle{definition}
\newtheorem*{definition}{Definition}

\theoremstyle{remark}
\newtheorem*{remark}{Remark}

\newtheorem{lemma}{Lemma}

\usepackage{listings}
\usepackage[usenames,dvipsnames]{color}
\definecolor{MyDarkGreen}{rgb}{0.0,0.4,0.0}
\begin{document}

\title{Hidden attractors on one path:\\
Glukhovsky-Dolzhansky, Lorenz, and Rabinovich systems
%from Glukhovsky-Dolzhansky system through Lorenz system to Rabinovich system
}

\author{G. Chen}
\affiliation{City University of Hong Kong, Hong Kong SAR, China}
\author{N.V. Kuznetsov}
%\email[]{Corresponding author: nkuznetsov239@gmail.com}
\affiliation{Faculty of Mathematics and Mechanics, St. Petersburg State University,
Peterhof, St. Petersburg, Russia}
\affiliation{Department of Mathematical Information Technology,
University of Jyv\"{a}skyl\"{a}, Jyv\"{a}skyl\"{a}, Finland}
\author{G.A. Leonov}
\affiliation{Faculty of Mathematics and Mechanics, St. Petersburg State University,
Peterhof, St. Petersburg, Russia}
\affiliation{Institute of Problems of Mechanical Engineering RAS, Russia}
\author{T.N. Mokaev}
\affiliation{Faculty of Mathematics and Mechanics, St. Petersburg State University,
Peterhof, St. Petersburg, Russia}

\date{\today}

\begin{abstract}
In this report, by the numerical continuation method we visualize and connect
hidden chaotic sets in the Glukhovsky-Dolzhansky, Lorenz and Rabinovich systems
using a certain path in the parameter space of a Lorenz-like system.
\end{abstract}

\maketitle

\section{\label{sec:intro} Introduction}

In 1963, meteorologist Edward Lorenz suggested an approximate
mathematical model (the \emph{Lorenz system})
for the Rayleigh-B\'{e}nard convection and
discovered numerically a chaotic attractor in this model \cite{Lorenz-1963}.
This discovery stimulated rapid development of
the chaos theory, numerical methods for attractor investigation, and
till now has received a great deal of attention from different fields
\cite{Celikovsky-1994,Li-2015-PhysLetA,Doedel-2015,LeonovK-2015-AMC,LeonovKKK-2016-CNSCS,Cermak-2017}.
The Lorenz system gave rise to various generalizations, e.g. {\it Lorenz-like systems},
some of which are also simplified mathematical models of physical phenomena.
In this paper, we consider the following Lorenz-like system
\begin{equation}
\begin{cases}
 \dot{x} $ = $ - \sigma(x - y) - a y z\\
 \dot{y} $ = $ r x - y - x z \\
 \dot{z} $ = $ -b z + x y,
\end{cases}
\label{sys:lorenz-general}
\end{equation}
where parameters $r$, $\sigma$, $b$ are positive and $a$ is real.
System \eqref{sys:lorenz-general} with
\begin{equation}\label{cond:lorenz}
  a = 0
\end{equation}
coincides with the classical Lorenz system.

Consider
\begin{equation}\label{cond:gd}
  b = 1, \qquad a > 0, \qquad \sigma > ar.
\end{equation}
 Then by the following linear transformation
(see, e.g., \cite{LeonovB-1992}):
\begin{equation}
  (x, y, z)  \rightarrow
  \bigg(x, \frac{\zeta}{\sigma - ar}, r - \frac{\zeta}{\sigma - ar} \, y\bigg),
\label{sys:lorenz-general:change_var}
\end{equation}
system \eqref{sys:lorenz-general}
is transformed to the {\it Glukhovsky-Dolzhansky system} \cite{GlukhovskyD-1980}:
\begin{equation}
\begin{cases}
 \dot{x} $ = $ -\sigma x + \zeta z + \alpha y z \\
 \dot{y} $ = $  \rho - y - xz \\
 \dot{z} $ = $ -z + xy,
\end{cases}
\label{sys:conv_fluid}
\end{equation}
where
\begin{equation}
 \zeta > 0, \quad
 \rho = \frac{r (\sigma - ar)}{\zeta} > 0, \quad
 \alpha = \frac{\zeta^2 a}{(\sigma - ar)^2} > 0.
 \label{sys:conv_fluid:param}
\end{equation}
The Glukhovsky-Dolzhansky system describes the convective fluid motion inside a rotating ellipsoidal cavity.

If we set
\begin{equation}\label{cond:rabinovich}
  a < 0, \qquad \sigma = -ar,
\end{equation}
then after the linear transformation (see, e.g., \cite{LeonovB-1992}):
\[
  (x, y, z)  \rightarrow
  \bigg(\nu_1^{-1} \, y, \nu_1^{-1} \nu_2^{-1} h \, x,  \nu_1^{-1} \nu_2^{-1} h \, z)
  \bigg), \quad
  t \to \nu_1 \, t
\]
with positive $\nu_1,\nu_2,h$,
we obtain the {\it Rabinovich system} \cite{Rabinovich-1978,PikovskiRT-1978},
describing the interaction of three resonantly coupled waves, two of which being
parametrically excited:
\begin{equation}
\begin{cases}
	\dot{x} $ = $ h y - \nu_1 x - y z, \\
	\dot{y} $ = $ h x - \nu_2 y + x z, \\
	\dot{z} $ = $ - z + x y,
\end{cases}
\label{sys:rabinovich}
\end{equation}
% Here the parameter $h$ is proportional to the pumping amplitude and the parameters
% $\nu_1$ and $\nu_2$ are normalized dumping decrements.
where
\begin{equation}
	\sigma = \nu_1^{-1} \nu_2,\,
	b = \nu_1^{-1},\,
	a = -\nu_2^2 h^{-2},\,
	r = \nu_1^{-1} \nu_2^{-1}h ^{2}.
	\label{eq:params-relation}
\end{equation}

Hereinafter, the Lorenz, Glukhovsky-Dolzhansky, and Rabinovich systems are studied in the
framework of system \eqref{sys:lorenz-general} under the corresponding
assumptions on parameters (\eqref{cond:lorenz}, \eqref{cond:gd}, or \eqref{cond:rabinovich}), respectively.
For the considered assumptions on parameters, if $r < 1$, then \eqref{sys:lorenz-general}
has a unique\footnote{
  In general, system \eqref{sys:lorenz-general} can possess up to five equilibria \cite{LeonovB-1992}.
} equilibrium ${\bf \rm S_0} = (0,0,0)$, which is
globally asymptotically Lyapunov stable
\cite{LeonovB-1992,BoichenkoLR-2005}.
If $r > 1$, then system \eqref{sys:lorenz-general}
has three equilibria: ${\bf \rm S_0} = (0,0,0)$ and
\begin{equation}
 {\bf \rm S_{\pm}} = (\pm x_1, \, \pm y_1, \, z_1). \label{eq:equil_s12}
\end{equation}
Here,
\[
 x_1 = \frac{\sigma b \sqrt{\xi}}{\sigma b + a \xi}, \quad
 y_1 = \sqrt{\xi}, \quad
 z_1 = \frac{\sigma \xi}{\sigma b + a \xi},
\]
and
\[
 \xi = \frac{\sigma b}{2 a^2} \left[ a (r-2) - \sigma + \sqrt{(\sigma - ar)^2 + 4a\sigma} \right].
\]
The stability of equilibria $S_{\pm}$ of system \eqref{sys:lorenz-general} depends on
the parameters $r$, $\sigma$, $a$ and $b$.

\begin{lemma}[see, e.g. \cite{LeonovKM-2015-EPJST}]\label{lemma:stability:gd}
  For a certain $\sigma > 2$, the equilibria $S_{\pm}$
  of system \eqref{sys:lorenz-general} with \eqref{cond:gd}
  (and, thus, of Glukhovsky-Dolzhansky system \eqref{sys:conv_fluid})
  are stable if and only if the following condition holds:
  \begin{equation}\label{eq:stability:gd}
    \begin{aligned}
      a^2 \sigma^2 (\sigma -2) r^3 &- a \left(2\sigma^4 - 4\sigma^3 - 3 a
     \sigma^2 + 4 a \sigma + 4 a\right) r^2 + & \\
      &+ \sigma^2 \left(\sigma^3 + 2(3 a - 1)\sigma^2 -
      8 a \sigma + 8 a\right) r - &\\
      &-\sigma^3 \left(\sigma^3 + 4\sigma^2
     - 16a\right) < 0. &
    \end{aligned}
  \end{equation}
\end{lemma}
\begin{lemma}[see, e.g. \cite{KuznetsovLMS-2016-INCAAM}]\label{lemma:stability:rabinovich}
The equilibria $S_{\pm}$ of
system \eqref{sys:lorenz-general} with \eqref{cond:rabinovich}
(and, thus, of the Rabinovich system \eqref{sys:rabinovich})
are stable if and only if one of the following conditions holds:
  \begin{enumerate}[label=(\roman*)]
    \item $0 \leq \, ar + 1 \, < \frac{2 r}{r - \sqrt{r(r-1)}}$, \label{cond:stability:1}
    \item $ar + 1 < 0$, \, $b > b_{\rm cr} =
    \frac{4 a (r - 1) (ar + 1) \sqrt{r(r-1)} + (ar - 1)^3}{(ar + 1)^2 - 4ar^2}.$
    \label{cond:stability:2}
  \end{enumerate}
\end{lemma}

% Note that system \eqref{sys:lorenz-general}
% is dissipative in the sense that it possesses bounded convex absorbing set
% $\mathcal{B} = \left\{(x,y,z) \in \mathbb{R}^3 ~|~ V(x,y,z) \leq \frac{b (\sigma + \delta r)^2}{2 c (a + \delta)} \right\}$,
% where $V(x,y,z) = x^2 + \delta y^2 + (a + \delta)\left(z - \frac{\sigma + \delta r}{a + \delta}\right)^2$
% is a Lyapunov function, $\delta$ is an arbitrary positive number such that
% $a + \delta > 0$ and $c = \min(\sigma, 1, \frac{b}{2})$ \cite{LeonovB-1992,LeonovKM-2015-EPJST}.
% Thus, system \eqref{sys:lorenz-general} generates a dynamical system and
% possesses a {\it global attractor} \cite{Chueshov-2002-book,LeonovKM-2015-EPJST}.

%The trivial attractors are stable equilibria and stable limit cycles.
The particular interest in the considered Lorenz-like systems is due to the
existence of chaotic attractors in their phase spaces.
In the next section, we will present the definition of attractor from
analytical and numerical perspectives.

\section{\label{sec:attractor} Attractors of dynamical systems}

\subsection{\label{subsec:attractor:analytial} Attractors of dynamical systems}

Consider system \eqref{sys:lorenz-general} as
an autonomous differential equation in a general form:
\begin{equation}\label{sys:ode}
  \dot{u} = f({u}), \qquad
\end{equation}
where $u=(x,y,z) \in \mathbb{R}^3$, and the
continuously differentiable vector-function $f: \mathbb{R}^3 \to \mathbb{R}^3$
may represent the right-hand side of system \eqref{sys:lorenz-general}.
Define by ${u}(t,{u}_0)$ a solution of \eqref{sys:ode} such that
${u}(0,{u}_0)={u}_0$.
%,and consider the evolutionary operator $\varphi^t({u}_0) = {u}(t,{u}_0)$.
For system \eqref{sys:ode}, a bounded closed invariant set K is
\begin{enumerate}[label=(\roman*)]
  \item a {\it (local)  attractor} if it is a minimal locally attractive set
        (i.e., $\lim_{t \to +\infty} {\rm dist} (K, {u}(t,{u}_0)) = 0$ for all
        ${u_0} \in K(\varepsilon)$, where $K(\varepsilon)$ is a
        certain $\varepsilon$-neighborhood of set $K$),
  \item a {\it global attractor} if it is a minimal globally attractive set
        (i.e., $\lim_{t \to +\infty} {\rm dist} (K, {u}(t,{u}_0)) = 0$
        for all ${u_0} \in \mathbb{R}^n$),
\end{enumerate}
where ${\rm dist}(K, {u}) = \inf_{{v} \in K} ||{v} - {u}||$
is the distance from the point ${u} \in \mathbb{R}^3$ to the set $K \subset \mathbb{R}^3$ (see, e.g. \cite{LeonovKM-2015-EPJST}).

Note that system \eqref{sys:lorenz-general} %(or \eqref{sys:ode})
is dissipative in the sense that it possesses a bounded convex absorbing set
\cite{LeonovB-1992,LeonovKM-2015-EPJST}
\begin{equation}\label{absorb_set}
  \mathcal{B} = \left\{(x,y,z) \in \mathbb{R}^3 ~|~ V(x,y,z) \leq \frac{b (\sigma + \delta r)^2}{2 c (a + \delta)} \right\},
\end{equation}
where $V(u)=V(x,y,z) = x^2 + \delta y^2 + (a + \delta)\left(z - \frac{\sigma + \delta r}{a + \delta}\right)^2$,
$\delta$ is an arbitrary positive number such that $a + \delta > 0$ and $c = \min(\sigma, 1, \frac{b}{2})$.
Thus, solutions of \eqref{sys:ode} exist for $t \in [0,+\infty)$
and system \eqref{sys:lorenz-general}
possesses a global attractor \cite{Chueshov-2002-book,LeonovKM-2015-EPJST},
which contains the set of equilibria
and can be constructed as
$\cap_{\tau > 0} \overline{\cup_{t \geq \tau} \varphi^t\left(\mathcal{B}\right)}$.
%From this and the uniqueness of solutions it follows that system \eqref{sys:ode}
%generates a \emph{dynamical system} $\{\varphi^t\}_{t\geq0}$.

%The above definitions of attractors
%implies that a global B-attractor is also a global attractor.
%Consequently, it is rational to introduce the notion of a
%{\it minimal global attractor} (or {\it minimal attractor})
%\cite{Chueshov-1993,Chueshov-2002-book}.
%This is the smallest bounded closed invariant set that
%is globally (locally) attractive.
%Further, the attractors (global attractors)
%will be interpreted as minimal attractors (minimal global attractors).

Computational errors (caused by finite precision arithmetic and numerical integration
of differential equations) and sensitivity to initial conditions %on chaotic attractor
allow one to get a reliable visualization of a chaotic attractor
by only one pseudo-trajectory computed on a sufficiently large time interval.
For that, one needs to choose an initial point in attractor's basin of attraction and
observe how the trajectory starting from this initial point
after a transient process visualizes the attractor.
Thus, from a computational point of view, it is natural
to suggest the following classification of attractors,
based on the simplicity of finding the basin of attraction in the phase space.

% \subsection{\label{subsec:attractor:numerical} Attractors from the computational perspective}
\begin{definition}{\cite{KuznetsovLV-2010-IFAC,LeonovKV-2011-PLA,LeonovK-2013-IJBC,LeonovKM-2015-EPJST}}
 An attractor is called a \emph{self-excited attractor}
 if its basin of attraction
 intersects with any open neighborhood of a stationary state (an equilibrium);
 %a small neighborhood of an equilibrium,
 otherwise, it is called a \emph{hidden attractor}.
\end{definition}

\begin{remark}
\emph{Sustained chaos} is often (almost) indistinguishable numerically
from \emph{transient chaos} (transient chaotic set in the phase space),
which can nevertheless persist for a long time.
Similar to the above definition, in general, a \emph{chaotic set} can be called \emph{hidden}
if it does not involve and attract trajectories from a small vicinities of stationary states;
otherwise, it is called \emph{self-excited}.
\end{remark}

For a \emph{self-excited attractor}, its basin of attraction
is connected with an unstable equilibrium
and, therefore, self-excited attractors
can be localized numerically by the
\emph{standard computational procedure}
in which after a transient process a trajectory,
started in a neighborhood of an unstable equilibrium (e.g., from a point of its unstable manifold),
is attracted to the state of oscillation and then traces it.
Thus, self-excited attractors can be easily visualized
(see, e.g. the classical Lorenz, Rossler, and Hennon  attractors can be visualized
by a trajectory from a vicinity of unstable zero equilibrium).
%The term \emph{self-excited oscillation} or {\it self-oscillation}
%can be traced back to the works of H.~Barkhausen and A.~Andronov,
%where it describes the generation and maintenance of a periodic motion
%in an electromechanical model by a source of power that lacks any corresponding periodicity
%(e.g., a stable limit cycle in the van der Pol oscillator) \citep{AndronovVKh-1966,Jenkins-2013}.

For a hidden attractor, its basin of attraction is not connected with equilibria, and,
thus, the search and visualization of hidden attractors in the phase space may be a challenging task.
Hidden attractors are attractors in the systems without equilibria
(see, e.g. rotating electromechanical systems with Sommerfeld effect
described in 1902 \cite{Sommerfeld-1902,KiselevaKL-2016-IFAC}),
and in the systems with only one stable equilibrium
(see, e.g. counterexamples \cite{LeonovK-2011-DAN,LeonovK-2013-IJBC}
to the Aizerman's (1949) and Kalman's (1957) conjectures
on the monostability of  nonlinear control systems
\cite{Aizerman-1949,Kalman-1957}).
One of the first related problems is the second part of Hilbert's 16th problem (1900) \cite{Hilbert-1901}
on the number and mutual disposition of limit cycles
in two-dimensional polynomial systems
where nested limit cycles (a special case of multistability and coexistence of attractors)
exhibit hidden periodic oscillations (see, e.g., \cite{Bautin-1939,KuznetsovKL-2013-DEDS,LeonovK-2013-IJBC}).
The \emph{classification of attractors as being hidden or self-excited}
was introduced by G.~Leonov and N.~Kuznetsov
in connection with the discovery of the first hidden Chua attractor \cite{LeonovK-2009-PhysCon,KuznetsovLV-2010-IFAC,LeonovKV-2011-PLA,LeonovKV-2012-PhysD,KuznetsovKLV-2013}
and has captured much attention of scientists from around the world
(see, e.g. \cite{LiZY-2014-HA,BurkinK-2014-HA,LiSprott-2014-HA,Chen-2015-IFAC-HA,ZhusubaliyevMCM-2015-HA,KuznetsovKMS-2015-HA,ChenLYBXW-2015-HA,Semenov20151553,PhamRFF-2014-HA,BorahR-2017-HA,MenacerLC-2016-HA,MessiasR-2017,Zelinka-2016-HA,DancaKC-2016,Danca-2016-HA,WeiPKW-2016-HA,PhamVJVK-2016-HA,JafariPGMK-2016-HA,DudkowskiJKKLP-2016}).

\subsection{Hidden attractor localization via numerical continuation method} % (fold)
One of the effective methods for numerical localization of hidden attractors
in multidimensional dynamical systems is based on the
{\it homotopy} and {\it numerical continuation method (NCM)}.
The idea is to construct a sequence of
similar systems such that for the first (starting) system
the initial point for numerical computation of oscillating
solution (starting oscillation)
can be obtained analytically.
Thus, it is often possible to consider the starting
system with self-excited starting oscillation;
then the transformation of this starting oscillation
in the phase space
is tracked numerically while passing from one system to another;
the last system corresponds to the system
in which a hidden attractor is searched.

For studying the scenario of transition to chaos,
we consider system \eqref{sys:ode} with $f(u) = f(u, \lambda)$,
where $\lambda \in \Lambda \subset \mathbb{R}^d$
is a vector of parameters, whose variation in the parameter space $\Lambda$
determines the scenario.
Let $\lambda_{\rm end} \in \Lambda$ define a point corresponding to the system,
where a hidden attractor is searched.
Choose a point $\lambda_{\rm begin} \in \Lambda$ such that
we can analytically or numerically localize a certain nontrivial (oscillating) attractor
$\mathcal{A}^1$ in system \eqref{sys:ode} with $\lambda = \lambda_{\rm begin}$
(e.g., one can consider an initial self-excited attractor defined by a trajectory ${u}^1(t)$
numerically integrated on a sufficiently large time interval $t \in [0, T]$
with initial point ${u}^1(0)$ in the vicinity of an unstable equilibrium).
Consider a {\it path}\footnote{
  In the simplest case, when $d = 1$, the path is a line segment.
} in the parameter space $\Lambda$ , i.e. a continuous function
$\gamma~:~ [0,\,1] \to \Lambda$, for which $\gamma(0) = \lambda_{\rm begin}$ and
$\gamma(1) = \lambda_{\rm end}$,
and a sequence of points $\{\lambda^j\}_{j=1}^k$ on the path,
where $\lambda^1 = \lambda_{\rm begin}$,
$\lambda^k = \lambda_{\rm end}$,
such that the distance between
$\lambda^j$ and $\lambda^{j+1}$
is sufficiently small.
On each next step of the procedure,
the initial point for a trajectory to be integrated
is chosen as the last point of the trajectory integrated on the previous step:
${u}^{j+1}(0) = {u}^{j}(T)$.
Following this procedure and sequentially increasing $j$,
two alternatives are possible:
the points of $\mathcal{A}^j$ are in the basin of attraction
of attractor $\mathcal{A}^{j+1}$,
or while passing from system \eqref{sys:ode} with $\lambda = \lambda^j$
to system \eqref{sys:ode} with $\lambda = \lambda^{j+1}$,
a loss of stability bifurcation is observed and attractor $\mathcal{A}^j$ vanishes.
If, while changing $\lambda$ from $\lambda_{\rm begin}$ to $\lambda_{\rm end}$,
there is no loss of stability bifurcation of the considered attractors,
then a hidden attractor for $\lambda^k = \lambda_{\rm end}$ (at the end of the procedure)
is localized.

Classical attractors obtained in the Lorenz, Rabinovich, and Glukhovsky-Dolzhansky systems
are self-excited, each can be visualized easily by a trajectory from a small vicinity
of one of the unstable equilibria
(see \cite{Lorenz-1963}, \cite{Rabinovich-1978}, \cite{GlukhovskyD-1980}, respectively).

Recently, hidden attractors were discovered in the
Glukhovsky-Dolzhansky system \eqref{sys:conv_fluid} for $\sigma = 4$
(see \cite{LeonovKM-2015-CNSNS,LeonovKM-2015-EPJST})
and in the Rabinovich system (see \cite{KuznetsovLMS-2016-INCAAM})
by numerical continuation method.
For $\sigma = 10$, $b = 8/3$ and $r = 24$ in the Lorenz system, there
exists a hidden bounded chaotic set
(similar  to the classical Lorenz attractor),
which is numerically indistinguishable
from sustained chaos
since it persists for a very long time
(see corresponding discussions in \cite{YorkeY-1979,YuanYW-2017-HA}).
Our aim here is to find a continuous path in the parameter space of system \eqref{sys:lorenz-general}
that connects the above hidden chaotic set in the
Lorenz system
with the hidden Glukhovsky-Dolzhansky and Rabinovich attractors.

\begin{figure}[!hb]
  \centering
  \includegraphics[width=0.45\textwidth]{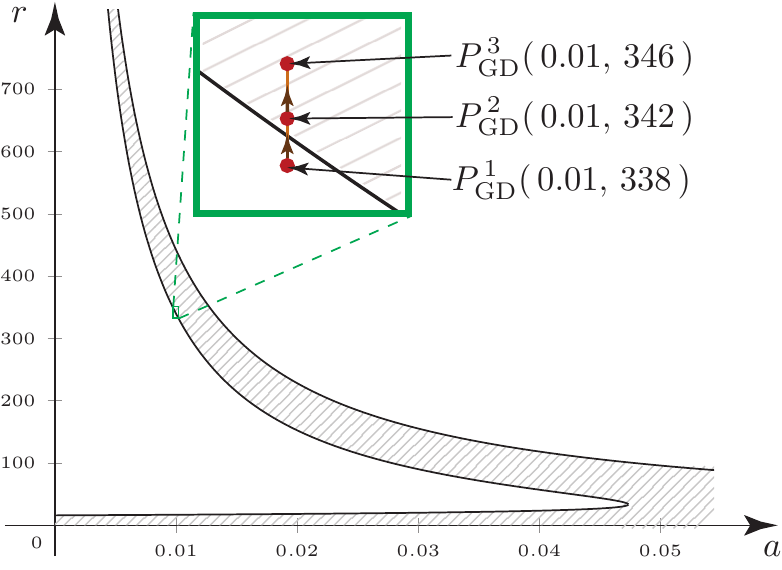}
  \caption{\label{fig:GD_path} Path $P_{\rm GD}^{\, 1} \to P_{\rm GD}^{\, 2} \to P_{\rm GD}^{\, 3}$
  in parameters plane $(a,r)$
  for localization of hidden GD attractor, $\sigma = 4$;\\
  $(\bullet)$ $P_{\rm GD}^{\, 1} (0.01, 338)$ : self-excited attractor with respect to $S_{0}$, $S_{\pm}$;
  $(\bullet)$ $P_{\rm GD}^{\, 2} (0.01, 342)$ : self-excited attractor with respect to $S_{0}$;
  $(\bullet)$ $P_{\rm GD}^{\, 3} (0.01, 346)$ : hidden attractor.
  Stability domain is defined according to inequality~\eqref{eq:stability:gd}.
  }
\end{figure}

\begin{figure*}[t]
 \centering
%\begin{minipage}{\textwidth}
 \includegraphics[width=0.95\textwidth]{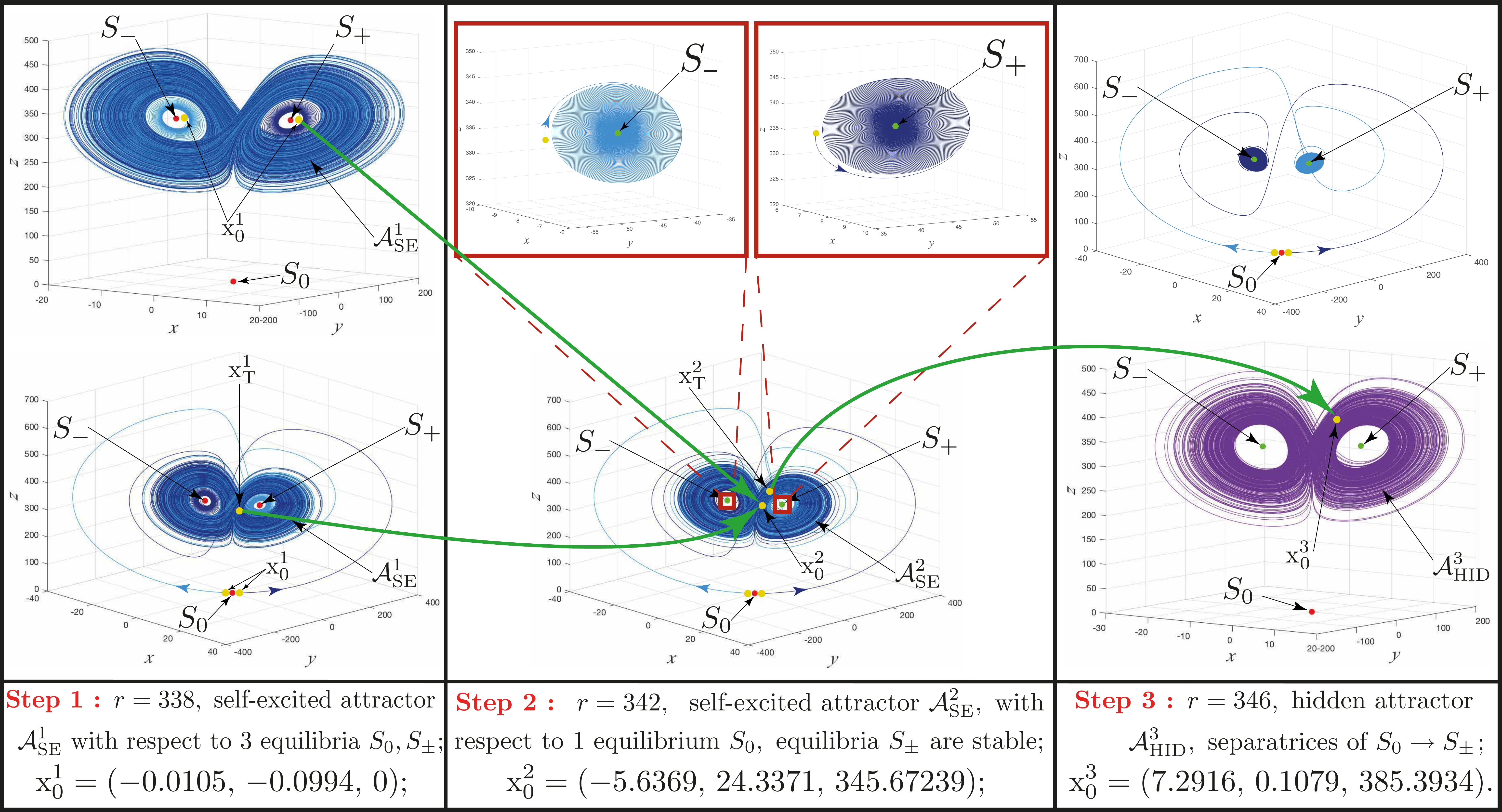}
 \caption{\label{fig:continuation_GD}
 Localization of hidden chaotic attractor (purple) in Glukhovsky-Dolzhansky systems
 defined by equations \eqref{sys:lorenz-general} and \eqref{cond:gd}
 using numerical continuation method.
 Here, trajectories ${\rm x}^i(t) = (x^i(t), y^i(t), z^i(t)$ (blue)
 are defined on the time interval
 $[0, T]$ ($T = 10^1$) and initial point on the $(i+1)$-th
 iteration (yellow) is defined as
 ${\rm x}_0^{i+1} := {\rm x}_T^{i}$ (light green arrows),
 where ${\rm x}_T^{i} = {\rm x}^{i}(T)$ is the final point (yellow).
 Outgoing separatrices of unstable zero equilibrium tend to two symmetric
 stable equilibria.}
%\end{minipage}
\end{figure*}

%\begin{minipage}{\textwidth}
\begin{figure*}[t]
 \includegraphics[width=0.95\textwidth]{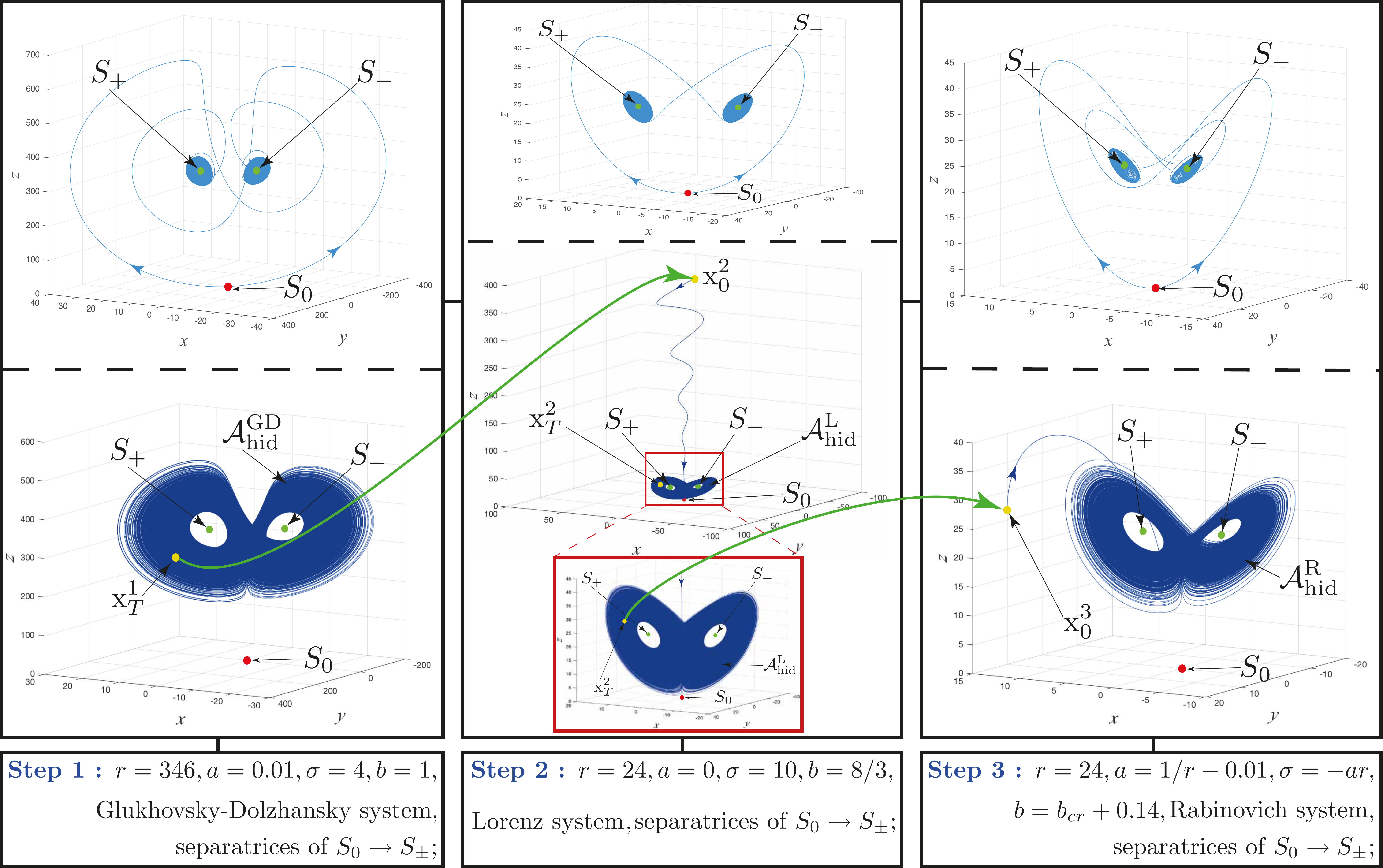}
 \caption{\label{fig:continuation}
 Localization of hidden chaotic sets in Glukhovsky-Dolzhansky, Lorenz and Rabinovich systems
 defined by equation \eqref{sys:lorenz-general} using numerical continuation method.
 Here, trajectories ${\rm x}^i(t) = (x^i(t), y^i(t), z^i(t)$ (blue)
 are defined on the time interval $[0, T]$,
 (${\rm GD} \to {\rm L}$ : $T = 10^4$; ${\rm L} \to {\rm R}$ : $T = 1.1 \cdot 10^4$)
 and initial point on the $(i+1)$-th iteration (yellow) is defined as
 ${\rm x}_0^{i+1} := {\rm x}_T^{i}$ (light green arrows),
 where ${\rm x}_T^{i} = {\rm x}^{i}(T)$ is the final point (yellow).
 Outgoing separatrices of unstable zero equilibrium tend to two symmetric
 stable equilibria.
 }
 %\end{minipage}
\end{figure*}

\section{\label{sec:attractor:hidden}
Localization of hidden attractors on one path}

In this experiment for system \eqref{sys:lorenz-general},
we consider three sets of parameters:
$P_{\rm GD} \,\left(r = 346, \, a = 0.01, \, \sigma = 4, \, b = 1\right)$
(for the Glukhovsky-Dolzhansky system --- GD),
$P_{\rm L}\, \left(r = 24, \, a = 0, \, \sigma = 10, \, b = 8/3\right)$
(for the Lorenz system --- L),
and $P_{\rm R} \,\left(r = 24, \, a = -1 / r - 0.01, \, \sigma = -ar, \, b = b_{\rm cr} + 0.14 \right)$
(for the Rabinovich system --- R).
Here, we change the parameters
in such a way that hidden Glukhovsky-Dolzhansky and Rabinovich attractors are located
not too close to the unstable zero equilibrium so as
to avoid a situation that numerically integrated trajectory
persists for a long time and then
falls on an unstable manifold of the unstable zero equilibrium,
then leaves the transient chaotic set, and finally tends to one of the stable equilibria
(see e.g. the corresponding discussion on the Lorenz system in \cite{YorkeY-1979}).

Hidden chaotic attractor in the Glukhovsky-Dolzhansky system
with $P_{\rm GD} \,\left(r = 346, \, a = 0.01, \, \sigma = 4, \, b = 1\right)$
can be obtained from a self-excited attractor
by numerical continuation method
\cite{LeonovKM-2015-CNSNS,LeonovKM-2015-EPJST}.
See the corresponding path in the space of parameters in Fig.~\ref{fig:GD_path}
and the localization procedure in Fig.~\ref{fig:continuation_GD}.

These sets of parameters define three points,
$P_{\rm GD}$, $P_{\rm L}$ and $P_{\rm R}$,
in the 4D parameter space $(r, \, a, \, \sigma, \,b)$.
Consider two line segments,
$P_{\rm GD} \rightarrow P_{\rm L}$ and $P_{\rm L} \rightarrow P_{\rm R}$,
defining two parts of the path in the continuation procedure.
Choose the partition of the line segments into $N_{\rm st} = 10$ parts and define intermediate
values of parameters as follows: $P_{\rm GD \to L}^i = P_{\rm GD} + \frac{i}{N_{\rm st}} (P_{\rm L} - P_{\rm GD})$ and
$P_{\rm L \to R}^i = P_{\rm L} + \frac{i}{N_{\rm st}} (P_{\rm R} - P_{\rm L})$, where $i = 1, \ldots, N_{\rm st}$.
Initial points for trajectories of system \eqref{sys:lorenz-general}
that define hidden chaotic sets are presented in Table~\ref{table:hidden:init}.
At each iteration of the procedure,
a chaotic attractor (defined by the trajectory in the phase space of system
\eqref{sys:lorenz-general}) is computed.
The last computed point of the trajectory at the previous step is used as
the initial point for computation at the next step.
\begin{table}[!h]
\centering
\caption{Initial point $(x_0,\,y_0,\,z_0)$ and time interval $[0,~T]$ of
numerical integration for each part of the path.}
\begin{tabular}{
|>{\centering}m{3cm}<{\centering}|
>{\centering}m{3.5cm}<{\centering}|
>{\centering}m{1.6cm}<{\centering}|}
\hline
Path & $(x_0,\,y_0,\,z_0)$ & $T$
\tabularnewline\hline
${\rm GD \to L}$ & $(10.64, \,  60.78, \, 390)$ & $10^4$
\tabularnewline\hline
${\rm L \to R}$ & $(0.2,\, 0.2,\, 0.35)$ & $1.1 \cdot 10^4$
\tabularnewline\hline
\end{tabular}
\label{table:hidden:init}
\end{table}
By this procedure,
starting from the hidden Glukhovsky-Dolzhansky attractor
it is possible to localize numerically hidden chaotic sets in the Lorenz
and Rabinovich systems.
For the considered parameters, the trajectories, starting in small
neighborhoods of unstable zero equilibrium, are not attracted by the computed chaotic set,
and the outgoing separatrices of unstable zero equilibrium tend to two symmetric stable equilibria.
Thus, the computed chaotic sets are hidden according to the above classification.

%And vice versa, it is possible to obtain hidden Lorenz and GD attractors, successively,
%starting from hidden Rabinovich attractor.
\begin{remark}
The path and its partition are chosen such that
during the procedure the obtained intermediate attractors are self-excited
(equilibria $S_{\pm}$ are unstable) and the basin of attraction of the attractor
at the current step intersects with the attractor obtained on the previous step.
\end{remark}

Hereinafter, it is reasonable to try to increase the length of the step
(i.e. decrease the number of the steps) in the continuation procedure,
but we may face the situation where the basin of attraction of the current attractor
does not intersect the previous attractor, or intersects it only partially.
In this case, the result of the procedure depends on the time interval of the
numerical integration of the trajectory.

%We have tried to localize hidden Lorenz attractor from hidden Glukhovsky-Dolzhansky attractor
%and hidden Rabinovich attractor from hidden Lorenz attractor, i.e. without
%intermediate steps in the continuation procedure.
%We obtain the following results:
%hidden Lorenz attractor can be localized directly from the hidden Glukhovsky-Dolzhansky attractor
%using the same time interval $[0, \, 10^4]$;
%hidden Rabinovich attractor can be localized directly from the hidden Lorenz attractor,
%but due to the mentioned difficulties the time interval has to be changed to the following
%$[0, \, 1.1 \cdot 10^4]$ (see Fig.~\ref{fig:continuation}).

All numerical experiments were performed in MATLAB R2016b
using standard procedures for numerical ODE integration.

\section{Conclusion} %\label{sec:conclusion}
In this report, by means of the numerical continuation method we localize
hidden chaotic sets on one path:
from the Glukhovsky-Dolzhansky system through the Lorenz system to
the Rabinovich system.
This helps better understanding of hidden chaotic attractors
and their relationships.

\section*{Acknowledgement} \label{sec:conclusion}
This work was supported by the Russian Science Foundation (project 14-21-00041).

%\bibliographystyle{elsarticle-num}
%\bibliography{../../bib/bib_nk,../../bib/bib_leonov,../../bib/bib_full,../../bib/bib_pll,../../bib/genlorenz-bib}
%\bibliography{C:/Dropbox/bib/bib_nk,C:/Dropbox/bib/bib_leonov,C:/Dropbox/bib/bib_full,C:/Dropbox/bib/bib_pll,C:/Dropbox/bib/bib-2008-str-at,C:/Dropbox/bib/genlorenz-bib}

%\newpage
\onecolumngrid
\begin{lstlisting}[float=*,
caption={{Numerical visualization of chaotic sets in the Glukhovsky-Dolzhansky, Lorenz and Rabinovich systems. %is presented in Listing~\ref{lst:plot_attravtors}.
\bf plotAttractors.m} -- plot GD, Lorenz and Rabinovich hidden attractors.}, label=lst:plot_attravtors]
function plotAttractors
% GD parameters (r, a > 0, sigma > a*r, b = 1) and initial point:
  r_GD = 346; a_GD = 0.01; b_GD = 1; sigma_GD = 4; trajGD_0 = [10,   60,   390];
% Lorenz paramters (r, b, sigma > 0, a = 0) and initial point:
  r_L = 24; a_L = 0; b_L = 8/3; sigma_L = 10; trajL_0 = [0.2,   0.2,   0.35];
% Rabinovich parameters (r, b > 0, a < 0, sigma = -a * r) and initial point:
  r_R = 24; a_R = -1 / r_R - 0.01; b_cr = (4 * a_R * (r_R - 1) * (a_R * r_R + 1) * ...
      sqrt(r_R * (r_R - 1)) + (a_R * r_R - 1)^3) / ((a_R * r_R + 1)^2 - 4 * a_R * r_R^2);
  b_R = b_cr + 0.14; sigma_R = - a_R * r_R; trajR_0 = [-2.4,   3.6,   23.6];
% Integration time:
  tEnd = 1e4;
% Plot GD, Lorenz and Rabinovich attractors:
  figure(1); plot3d(trajGD_0, tEnd, r_GD, a_GD, b_GD, sigma_GD);
  figure(2); plot3d(trajL_0, tEnd, r_L, a_L, b_L, sigma_L);
  figure(3); plot3d(trajR_0, tEnd, r_R, a_R, b_R, sigma_R);
% Generalized Lorenz system:
    function out = genLorSys(t, x, r, a, b, sigma)
        out = zeros(3,1);
        out(1) = - sigma * x(1) + sigma * x(2) - a * x(2) * x(3);
        out(2) = r * x(1) - x(2) - x(1) * x(3);
        out(3) = x(1) * x(2) - b * x(3);
    end
% Attractor plotting routine:
    function plot3d(traj0, tEnd, r, a, b, sigma)
        % ODE solver settings:
        acc = 1e-8; RelTol = acc; AbsTol = acc; InitialStep = acc/10;
        solverOptions = odeset('RelTol', RelTol, 'AbsTol', AbsTol, ...
                          'InitialStep', InitialStep, 'NormControl', 'on');
        % Integration of the trajectory:
        [~, traj] = ode45(@(t, x) genLorSys(t, x, r, a, b, sigma), ...
                                          [0, tEnd], traj0, solverOptions);
        % Equilibria:
        S0 = [0 0 0];

        if a == 0
            S12XY = sqrt(b*(r-1)); S12Z = r-1;
            S1 = [S12XY, S12XY, S12Z]; S2 = [-S12XY, -S12XY, S12Z];
        else
            XSI = (sigma*b)/(2*a^2)*(a*(r-2)-sigma + sqrt((a*r-sigma)^2+4*a*sigma));
            S12X1 = (sigma*b*sqrt(XSI))/(sigma*b + a*XSI);
            S12Y1 = sqrt(XSI); S12Z1 = (sigma*XSI)/(sigma*b + a*XSI);
            S1 = [S12X1, S12Y1, S12Z1]; S2 = [-S12X1, -S12Y1, S12Z1];
        end

        plot3(S0(1), S0(2), S0(3), '.', 'markersize', 15, 'Color', 'red'); hold on;
        text(0, 0, 0,'S_0','fontsize', 18);

        plot3(S1(1), S1(2), S1(3), '.', 'markersize', 20, 'Color', 'green');
        text(S1(1), S1(2), S1(3),'S_1','fontsize', 18);

        plot3(S2(1), S2(2), S2(3), '.', 'markersize', 20, 'Color', 'green');
        text(S2(1), S2(2), S2(3),'S_2','fontsize', 18);

        plot3(traj(:, 1), traj(:, 2), traj(:, 3), 'Color', 'blue'); hold off;

        xlabel('x'); ylabel('y'); zlabel('z');
        grid on; axis auto; view(3);
    end
end
\end{lstlisting}

\end{document}